\pdfoutput=1
\input sgml.sty
\documentclass{146-mpla}
\usepackage{eso-pic}
\usepackage{color}

\definecolor{mygray}{gray}{.1}

\def\1streading{
  \AddToShipoutPicture{%
    \AtPageUpperLeft{%
      \makebox(350,40)[c]{\resizebox{\textwidth}{!}{%
        \rotatebox{0}{{\textbf{\color{mygray}{\hskip2.5in 1st Reading}}}}}}
    }
  }}

\def\2ndreading{
  \AddToShipoutPicture{%
    \AtPageUpperLeft{%
      \makebox(350,40)[c]{\resizebox{\textwidth}{!}{%
        \rotatebox{0}{{\textbf{\color{mygray}{\hskip2.5in 2nd Reading}}}}}}
    }
  }}

\newtoks\pii\pii{\jobname}
\pii{\jobname}
\def\doii{10.1142/\the\pii}
\def\fmdoi{\footnotesize DOI:~\href{http://dx.doi.org/\doii}{\doii}}

\usepackage[breaklinks]{hyperref}
\usepackage[hyphenbreaks]{breakurl}
\usepackage{xcolor}
\hypersetup{%
  colorlinks=false, 
  urlbordercolor=cyan,citebordercolor=cyan,linkbordercolor=cyan,
  pdfborderstyle={/S/U/W .5} 
}

\def\doi#1{doi:\href{http://dx.doi.org/#1}{\nolinkurl{#1}}}
\def\arxiv#1{arXiv:\href{https://arxiv.org/abs/#1}{\nolinkurl{#1}}}

\usepackage[mathlines,pagewise]{lineno}
\usepackage[super,sort,nospace,compress]{cite}
\def\refcite#1{\def\citepunct{, }\citen{#1}\def\citepunct{,}}
\def\myfig#1#2{\centerline{\includegraphics[#1]{\ArtDir/#2}}}

\makeatletter
\def\catchline#1#2#3#4#5{
 \clcount=#3
 \expandafter\def\expandafter\@clinebuf\expandafter
 {\@clinebuf\catchlinefont
\if@plogo
 \vspace*{-6pc}
 \hfill\hspace*{5.5pc}\includegraphics{\logoname}
 \hspace*{-5.7pc}
 \vspace*{6pc}
 \par
 \vspace*{-4.25pc}
\fi
 \vspace*{-.1pc}
 \noindent Modern Physics Letters A\par
 \noindent Vol.\ {#1}, \No\ {#2} (#3)\ \artid\ (\theeepage\ pages)\par
 \noindent \copyright\ The Author(s)\par
 \noindent\vskip-\baselineskip \hphantom{#4 \hskip2em #5}\par
 \noindent \fmdoi\par\vskip-11pt
 \vspace*{-28pt}
 \hbox to \hsize{\hfill\hspace*{12pt}{\includegraphics[width=3.3cm]{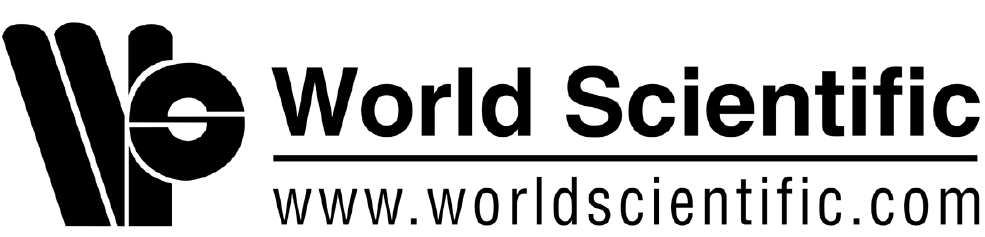}}}\par
 \vspace*{19pt}
 }\relax\par
 }
 \makeatother

\begin{document}

\def\artid{1650157}
\def\ArtDir{./}
\emid{MPLA-D-15-00298}


\setcounter{page}{1}

\markboth{E. T. Kipreos \& R. S. Balachandran}
{An approach to directly probe simultaneity}

\catchline{31}{26}{2016}{}{}

\title{An approach to directly probe simultaneity\footnotetext{This is an Open Access article published by World Scientific Publishing Company. It is distributed under the terms of the \href{https://creativecommons.org/licenses/by/4.0}{Creative Commons Attribution 4.0 (CC-BY) License}. Further distribution of this work is permitted, provided the original work is properly cited.}}

\author{Edward T. Kipreos$^*$ and Riju S. Balachandran$^\dag$}

\address{University of Georgia, 120 Cedar Street, Athens, GA 30602, USA\\
\email[*\!]{ekipreos@uga.edu}
\email[\dag\!]{rijub@uga.edu}}

\maketitle

\begin{history}
\received{28 September 2015}
\revised{15 April 2016}
\accepted{27 June 2016}
\published{2 August 2016}
\end{history}

\begin{abstract}
The theory of special relativity derives from the Lorentz transformation.  The Lorentz transformation implies differential simultaneity and light speed isotropy.  Experiments to probe differential simultaneity should be able to distinguish the Lorentz transformation from a kinematically-similar alternate transformation that predicts absolute simultaneity, the absolute Lorentz transformation.  Here, we describe how published optical tests of light speed isotropy/anisotropy cannot distinguish between the two transformations.  We show that the shared equations of the two transformations, from the perspective of the ``stationary'' observer, are sufficient to predict null results in optical resonator experiments and in tests of frequency changes in one-way light paths.  In an influential 1910 exposition on differential simultaneity, Comstock described how a ``stationary'' observer would observe different clock readings for spatially-separated ``moving'' clocks.  The difference in clock readings is an integral aspect of differential simultaneity.  We derive the equation for the difference in clock readings and show that it is equivalent to the Sagnac correction that describes light speed anisotropies in satellite communications.  We describe an experimental strategy that can measure the differences in spatially-separated clock times to allow a direct probe of the nature of simultaneity.
\end{abstract}

\keywords{Special relativity; Lorentz transformation; light speed isotropy.}

\ccode{PACS No.: 03.30.+p}

\enlargethispage*{13pt}

\section{Introduction}\label{sec1}
The theory of special relativity (SR) is based on the Lorentz transformation (LT), which describes how changes in velocity are linked to time dilation and length contraction.  The LT implies that length contraction and time dilation are reciprocally observed between two inertial reference frames (IRFs).  Embedded in the LT is differential simultaneity and the isotropic speed of light, denoted $c$.  In order to experimentally assess differential simultaneity, it is necessary to be able to distinguish the LT from alternate transformations that predict absolute simultaneity.  Here, we will show that current optical tests are unable to distinguish between the LT and a kinematically-similar transformation theory that predicts absolute simultaneity, absolute simultaneity theory (AST).  We will clarify an equation for a key attribute of differential simultaneity and describe an experimental strategy that measures this attribute to probe the nature of simultaneity.

\subsection{Comparison of the Lorentz transformation and ALT}\label{sec1.1}
The only alternate transformation that is compatible with classical tests of SR is the absolute Lorentz transformation (ALT).\cite{1,2}  ALT was named by Tangherlini, who described the transformation in 1958.\cite{3}  ALT was initially published by Eagle in 1938.\cite{4} In the modern era, Eagle's work does not appear to have been recognized as the first publication of the ALT equations until it was reported in two 2009 reviews.\cite{5,6}

The LT and ALT share the same length transformation equation:
\begin{equation}
x' = \frac{x - vt}{\sqrt{1-\frac{v^2}{c^2}}}\,.\label{1}
\end{equation}
The two transformations differ in their time transformation equations.  The LT time transformation equation is:
\noindent
\begin{equation}
t' = \frac{t - \frac{vx}{c^2}}{\sqrt {1-\frac{v^2}{c^2}}}\,.\label{2}
\end{equation}
The ALT time transformation equation from the stationary perspective is:
\begin{equation}
t' = t\sqrt {1-\frac{v^2}{c^2}}\,.\label{3}
\end{equation}
In the LT framework, $x'$ and $t'$ are ``moving'' coordinates for distance and time, respectively; and $x$ and $t$ are ``stationary'' coordinates.  ``Moving'' and ``stationary'' coordinate systems are interchangeable, and the velocity $(v)$ is calculated between two IRFs.  ALT implies the presence of a preferred reference frame (PRF) in which light speed is isotropic.  In the ALT framework, $x$ and $t$ are stationary PRF coordinates, and the velocity of objects is calculated relative to the PRF.

Experimental tests of time dilation in SR use Eq.~(\ref{3}), the ALT time transformation equation.  This is because experimental tests generally analyze movements in linear, constant-velocity IRFs for which $x = vt$.  Substituting $x = vt$ in the LT time transformation equation (\ref{2}) produces the ALT time transformation equation (\ref{3}).\cite{7}  Therefore, in experimental settings, the two theories calculate time dilation and length contraction using the same Eqs.~(\ref{1}) and (\ref{3}) if the measurements are taken from the ``stationary'' (PRF) reference frame.

\begin{figure}[t]
\vspace*{-2pt}
\hfil
\begin{minipage}[b]{5.1cm}
\myfig{width=5.1cm}{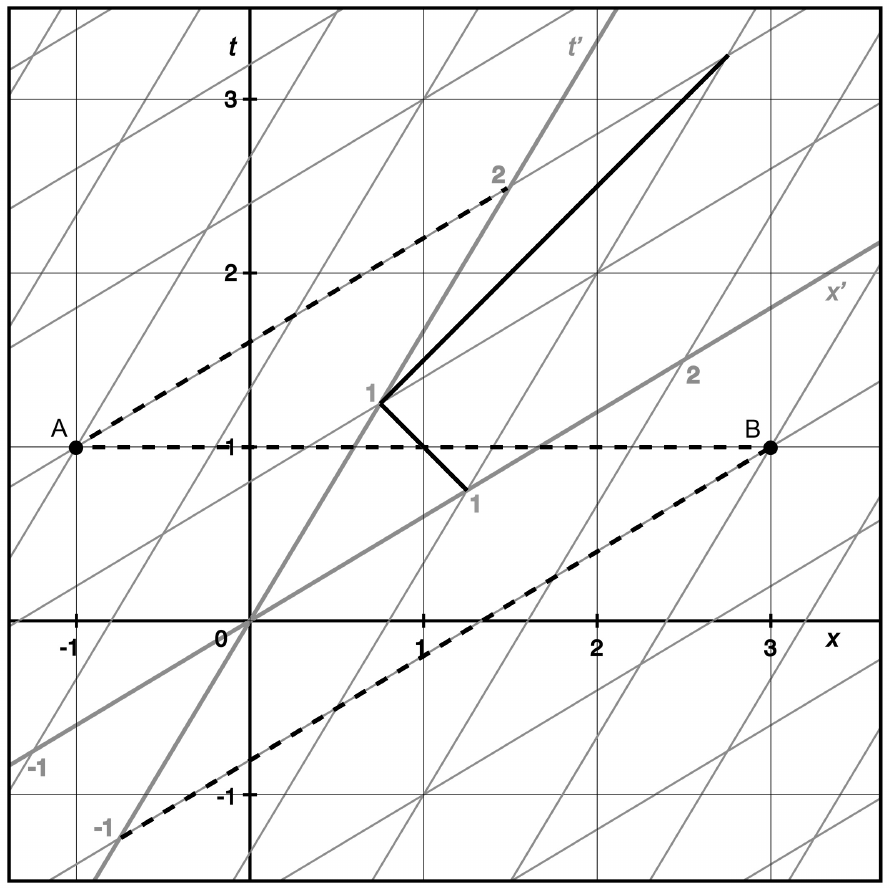}
\vspace*{4pt}
\centerline{\footnotesize (a)}
\end{minipage}
\hfil
\begin{minipage}[b]{5.1cm}
\myfig{width=5.1cm}{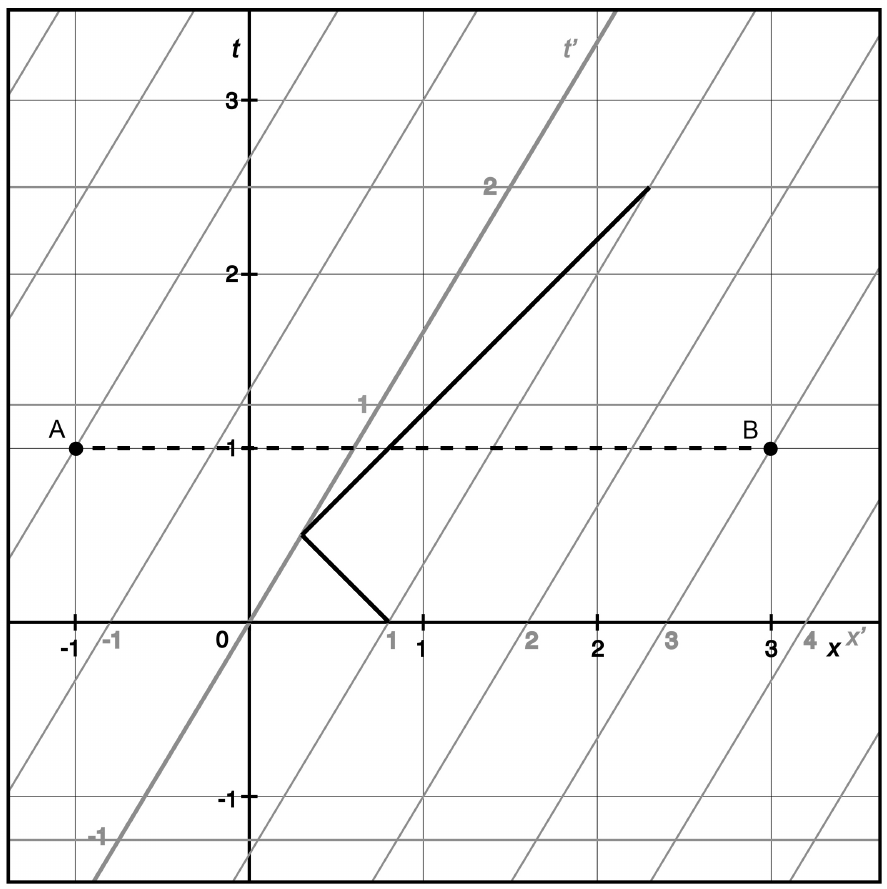}
\vspace*{4pt}
\centerline{\footnotesize (b)}
\end{minipage}
\caption{(a)~Minkowski and (b)~ALT spacetime diagrams for $v = 0.6c$.  Time units are femtoseconds (fs), and distance units are light-fs. Points (A) and (B) are events that occur simultaneously in the ``stationary'' frame at $t$ = 1 fs (dashed horizontal line).  In (a), (A) occurs at $t'=+2.0$ fs and (B) occurs at $t'=-1.0$~fs (dashed diagonal lines).  In (b), (A) and (B) occur simultaneously at $t'=+0.8$~fs.  To illustrate light speeds, light (solid black line) is sent at $t' = 0$ from $x' = 1$ towards $x' = 0$, where it is reflected back towards $x' = 1$.  In (a), the ``moving'' observer calculates isotropic light speeds.  In (b), the moving observer calculates one-way light speeds of $2.5c$ in the backward direction, and $0.625c$ in the forward direction; and a two-way light speed of $c$.}
\label{fig1}
\end{figure}

The ALT equations for the stationary and moving observers are not reciprocal,\cite{3} and the moving (non-PRF) observer will observe the opposite effects of time contraction and length extension when viewing clocks and objects in the PRF.  ALT therefore predicts directional and absolute time dilation and length contraction effects based on the object's movement relative to a PRF.  In ALT there is no theoretical justification for preferential treatment of IRFs, and ALT transformations apply equivalently to all motion relative to a PRF, both inertial and non-inertial.  In contrast, SR predicts reciprocal time dilation and length contraction between objects in IRFs based on the velocity between IRFs.  With ALT, one-way light speeds are isotropic in the PRF, but are anisotropic in other reference frames; the two-way speed of light is $c$ in all reference frames (Fig.~\ref{fig1}(b)).\cite{8}

\begin{figure}[t]
\vspace*{-3pt}
\hfil
\begin{minipage}[b]{5.1cm}
\myfig{width=5.1cm}{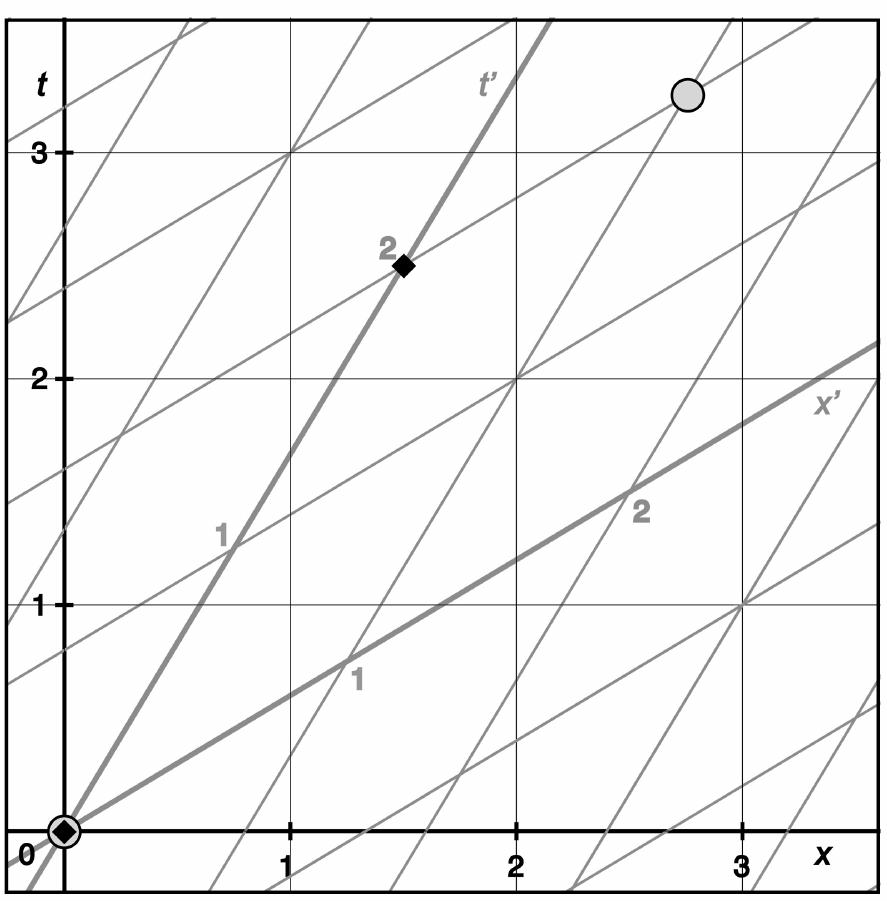}
\vspace*{4pt}
\centerline{\footnotesize (a)}
\end{minipage}
\hfil
\begin{minipage}[b]{5.1cm}
\myfig{width=5.1cm}{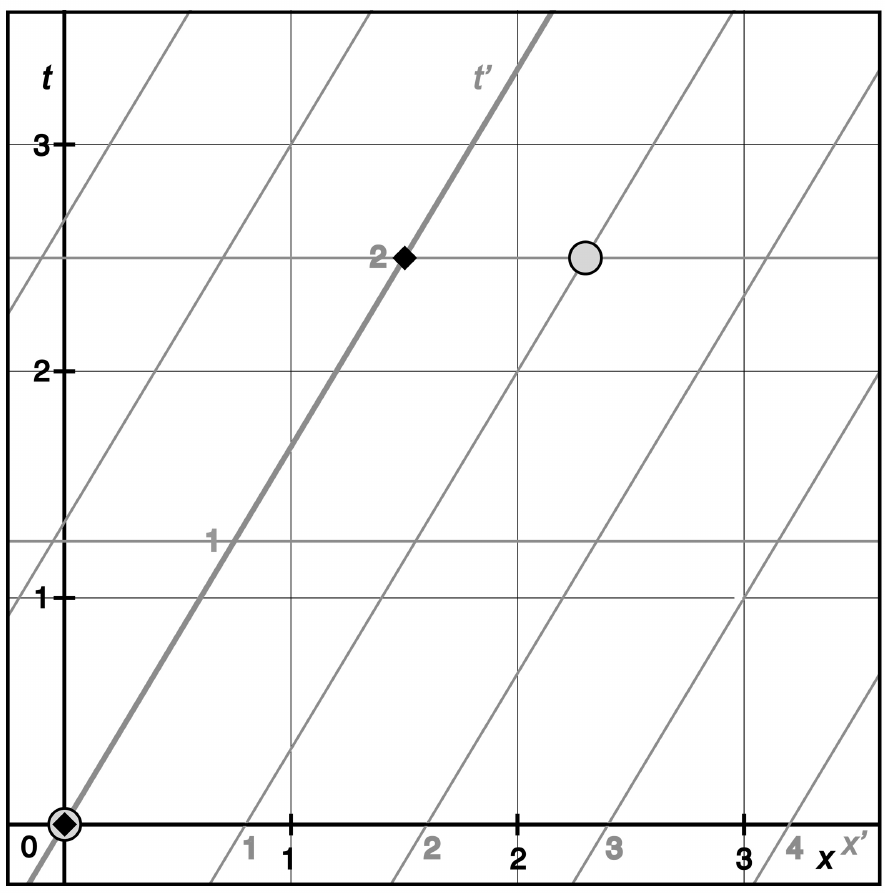}
\vspace*{4pt}
\centerline{\footnotesize (b)}
\end{minipage}
\caption{(a)~Minkowski and (b)~ALT spacetime diagrams for $v = 0.6c$ illustrating how the spatial separation of clocks introduces a RTO for SR but not for ALT.  For both diagrams, at $t' = 0$, two clocks (small diamond and large circle) are at $x' = 0$ light-fs.  At $t' = 2$~fs, two spatially-separated synchronized clocks are shown at positions $x' = 0$ and $x' = 1$.  In (a), a ``stationary'' intelligent observer can determine the RTO for the separated clocks at $t = 2.0$ by noting that the diamond clock reads $t' = 1.6$, and the circle clock reads
$t' = 1.0$, a difference of 0.6~fs.}
\label{fig2}
\end{figure}

SR is generally considered to be a unique solution to  the theory of general relativity (GR) with zero curvature.  Tangherlini reported that both ALT and SR are compatible with GR.\cite{3}  He showed that obtaining SR as a unique solution involves applying the restriction that all uniformly translating frames are equivalent in all respects.  In the absence of this restriction, ALT is also a solution to GR with zero curvature.\cite{3}

\subsection{In an absolute simultaneity framework$,$ ALT cannot be resynchronized to SR} \label{sec1.2}
\looseness-1It has been proposed that SR and ALT are equivalent because ALT coordinates can be resynchronized to Minkowski coordinates.\cite{9}  However, resynchronization would obscure the two theories, as ALT would then predict differential simultaneity rather than absolute simultaneity (Fig.~\ref{fig1}).\cite{1}  Further, it would not be possible to resynchronize the clocks properly in an absolute simultaneity framework.  The $-vx/c^2$ term in the LT time transformation equation (\ref{2}) indicates that different observers will see different times on a given clock based on their distance and velocity relative to the clock.  For resynchronization to occur in an absolute simultaneity framework, a clock would need to be resynchronized to simultaneously match the times expected by different observers.  However, absolute simultaneity implies that all observers will observe the same time on the clock at a given instant.  Therefore, within an absolute simultaneity framework, it would not be possible to resynchronize a clock to accommodate multiple moving observer's expectations for different times on the clock.

ALT is similarly incompatible with Einstein synchronization, which imposes isotropic one-way light speeds.\cite{7}  While Einstein synchronization is compatible with ALT if performed in the PRF, where light speed is isotropic, the implementation of Einstein synchronization in non-PRF moving reference frames would transform ALT coordinates into Minkowski coordinates.\cite{1}  Conversely, synchronizing clocks with instantaneous signals would transform Minkowski coordinates into ALT coordinates.\cite{3,5}  Therefore, experiments to distinguish differential simultaneity from absolute simultaneity cannot use Einstein synchronization or instantaneous synchronization.  Both theories are compatible with ``slow clock transport'',\cite{1}  in which clocks are synchronized in one location and then spatially-separated equivalent distances.  For both theories, the clocks maintain their synchrony within the ``moving'' reference frame, and the synchronization does not alter the two transformations' respective coordinate systems (Fig.~\ref{fig2}).

With ALT, velocities are calculated relative to PRFs.  Therefore, the choice of PRF affects the predicted time dilation and length contraction.  The cosmic micro\-wave background (CMB) radiation  has been considered as the PRF for ALT.\cite{1,10}  However, as the Earth is moving relative to the CMB by $\sim$\,368~km/s,\cite{11} that velocity would have to be added to velocities calculated on the Earth, which would make ALT predictions incompatible with multiple tests of SR.  It has been suggested that ALT can be resynchronized so that the isotropic ``privileged'' inertial system is transferred to another IRF.  The resynchronized IRF becomes the isotropic privileged reference frame, and all other reference frames are then not isotropic (including\- the original privileged reference frame).\cite{12}  It was suggested that the resynchronization would make it impossible to detect the true privileged reference frame.\cite{12}  In regards to this idea, we note that there is a lack of a plausible, real-world mechanism that would mediate the clock resynchronizations required to shift the PRF; additionally, the arbitrary nature of the PRFs would preclude testing ALT because PRFs could be assigned \textit{ex post facto}.

For the reasons described above, we think that the proper approach to consider the validity of ALT is in the absence of resynchronization.  Without resynchronization, ALT is only compatible with current experimental results in the situation that PRFs are locally associated with centers of gravitational mass.\cite{13}  This scenario provides velocities relative to the PRF that allow ALT to match experimental results for time dilation.\cite{13}  In this work, we will discuss how this type of PRF allows ALT to match current experimental data for light speed isotropy/anisotropy.

\subsection{AST encompasses ALT with PRFs that are locally associated with centers of gravitational mass}
\label{sec1.3}
The combination of ALT with gravity-centered PRFs has been termed AST.\cite{13}  The extent of AST-defined PRFs are potentially similar to the spheres of influence (SOIs) defined in the field of astrodynamics.  SOIs are mathematically defined regions around a celestial body where the primary gravitational influence is from the celestial body.\cite{14}  The patched conic approximation, based on SOIs, has been used to design interplanetary and lunar trajectories.\cite{14}  AST proposes that an object near a planet would have as its PRF the relevant planet-centered non-rotating IRF, similar to the SOI of the planet.  When leaving a planet-centered PRF, the object would enter the Sun's PRF, which is the heliocentric inertial (HCI) reference frame.  If AST were to be found to be valid, the extent to which PRFs approximate SOIs and the interactions between PRFs would need to be experimentally determined.

In the vicinity of the Earth, the relevant PRF for AST is the non-rotating Earth-centered inertial (ECI) reference frame.
Notably, the ECI is often used as the ``stationary'' reference frame for calculating relativistic transformations and certain quantum mechanical effects.\cite{15,16,17,18,19,20,21,22}  The ECI is used as the ``stationary'' reference frame for experimental tests of SR because it is an IRF.\cite{18}  Therefore, in many experimental settings, SR and AST utilize the same ``stationary'' reference frame with the same Eqs.~(\ref{1}) and (\ref{3}) to predict results, thereby producing equivalent predictions that preclude the experiments from distinguishing the two theories.

AST is an incomplete theory.  It is unclear how or why PRFs are linked to local centers of gravitational mass, or how PRFs interact.  This has implications for the velocity of objects relative to the PRF as they approach the boundaries of gravitational influence, e.g.~the Lagrangian point.  Despite this, it is worthwhile to experimentally distinguish AST from SR, because AST is the only potential alternate transformation theory that is compatible with current experiments.  If AST were invalidated by experimental results, then it would remove the last viable alternate to SR.

\section{Optical Experiments that are Incapable of Distinguishing SR and AST}\label{sec2}
\subsection{Two-way light path and optical resonator experiments}\label{sec2.1}
When considered from the perspective of a PRF observer, ALT shares a host of kinematics with SR, including that the velocity of light is independent of the velocity of its source,\cite{3} and kinematic equations for: length contraction (\ref{1}); time dilation for a constant-velocity IRF (\ref{3}); relativistic Doppler shift;\cite{8,23} and relativistic stellar aberration angle.\cite{8,23}

The shared kinematics of SR and ALT when considered from the ``stationary'' perspective are sufficient to produce null results in two-way light path experiments.  The null results of Michelson--Morley type experiments are explained by the effects of length contraction and the relativistic stellar aberration angle.\cite{1,24}  The null results of Kennedy--Thorndike type experiments are explained by the effects of length contraction, time dilation and relativistic stellar aberration angle.\cite{1,24}

Here, we will demonstrate that the shared kinematics of SR and ALT (from the ``stationary'' perspective) are sufficient to produce null results in optical resonator experiments.  These are two-way light path experiments that utilize orthogonally-arranged optical resonators.\cite{25,26,27,28,29,30,31,32,33,34,35,36,37}  The frequencies of light required to maintain standing optical waves in the two cavities are compared to determine if they change with motion.

To demonstrate that SR kinematic equations are sufficient to produce a null result, one demonstrates that a ``stationary'' observer will observe a null result based on the effects of the kinematics.  To demonstrate this for optical resonator experiments, consider orthogonal cavities in which one cavity is aligned in the direction of motion.  A standing wave requires that the cavity length~$(l)$ is a multiple of 1/2 the wavelength $(\lambda)$.  Therefore, to maintain the same number of standing waves per cavity length when in motion, the ratio of the ``moving'' wavelength $(\lambda')$ to the translatory distance that the light travels according to the ``stationary'' observer $(l_t')$ must maintain the ratio $\lambda/l$.

We will first consider the optical resonator parallel to the direction of motion.  The relativistic Doppler shift equation defines
$\lambda'$.  In the forward direction, the ``stationary'' observer observes a $l_t'$ of $cl'/(c-v)$; substituting
$l(1-v^2/c^2)^{0.5}$ for $l'$ produces $cl(1-v^2/c^2)^{0.5}/(c-v)$.  The ratios $\lambda'/l_t'$ and $\lambda/l$ are therefore equivalent:
\noindent
\begin{equation}
\frac{\lambda'}{l_t'}=\frac{\lambda\sqrt{\frac{1+\frac{v}{c}}{1-\frac{v}{c}}}}
{\frac{cl\sqrt{1-\frac{v^2}{c^2}}}{c-v}}=\frac{\lambda}{l}\,.\label{4}
\end{equation}
The ratios are also equivalent when light is sent in the backwards direction:
\begin{equation}
\frac{\lambda'}{l_t'}=\frac{\lambda\sqrt{\frac{1-\frac{v}{c}}{1+\frac{v}{c}}}}
{\frac{cl\sqrt{1-\frac{v^2}{c^2}}}{c+v}}=\frac{\lambda}{l}\,.\label{5}
\end{equation}

We will now consider the optical resonator aligned at a right angle to the direction of motion. The transverse light path $(l_t')$, as viewed by the ``stationary'' observer, can be calculated using the relativistic stellar aberration angle equation as
$l_t' = l/(1-v^2/c^2)^{0.5}$.  By definition, $\lambda = c/f$, where $f$ is the frequency of the light; and therefore $\lambda' = c/f'$.  Because of the effect of time dilation, $c/f' = c/f(1-v^2/c^2)^{0.5}$.  Substituting $\lambda$ for $c/f$ gives
$\lambda' = \lambda/(1-v^2/c^2)^{0.5}$.  The ratio $\lambda'/l_t'$ is therefore equivalent to $\lambda/l$:
\noindent
\begin{equation}
\frac{\lambda'}{l_t'}=\frac{\frac{\lambda}{\sqrt{1-\frac{v^2}{c^2}}}}
{\frac{l}{\sqrt{1-\frac{v^2}{c^2}}}}=\frac{\lambda}{l}\,.\label{6}
\end{equation}
Thus, the shared kinematics of length contraction, time dilation, relativistic Doppler shift and relativistic stellar aberration angle (from the ``stationary'' perspective) are sufficient to produce null results in optical resonator experiments.

\subsection{One-way light path experiments}\label{sec2.2}
Reviews have described five types of experiments that have been used to search for anisotropies in the speed of light using one-way light paths.\cite{38,39}  We will describe how these experiments are unable to distinguish between SR and AST.

Vessot {\it et~al.} conducted an experiment that compared the hydrogen maser clock frequencies sent between a rocket and Earth-surface locations, and calculated the one-way speed of light during the rocket's trajectory.\cite{19,20}  In this experiment,\break the ECI was used as the coordinate system.  AST predicts isotropic light speeds\break in the ECI, and therefore both theories predict isotropic light speeds.

Experiments that used the Compton scattering of laser photons on high-energy electrons to determine the one-way speed of light, analyzed changes in light speed as the Earth rotates and orbits the Sun.\cite{40,41,42}  The motion of the Earth alters the orientation of the experimental apparatus relative to a potential external PRF.  However, the direction and speed of the apparatus did not change relative to the ECI, and therefore no changes with time are expected in the AST framework.

Rotating M\"ossbauer experiments also analyze changes during Earth's rotational and orbital motion,\cite{43,44,45} but as there is no differential movement relative to the ECI, no changes with time are predicted in the AST framework.  Additionally, Ruderfer, who was the first to propose using rotating M\"ossbauer experiments as a test of SR,\cite{46} published an erratum that stated that theories that incorporate relativistic effects, such as ALT, would also predict null results.\cite{47,48}

Riis {\it et~al.} conducted an experiment analyzing the first-order Doppler shift of light emitted by an atomic beam using fast-beam laser spectroscopy.\cite{49}  This was subsequently shown to be incapable of distinguishing SR from ALT.\cite{50,51}

Krisher {\it et~al.} sought to determine changes in the one-way speed of light in response to the Earth's rotational and orbital motion by analyzing changes in maser frequencies after one-way travel over a 21~km fiber optics cable.\cite{52}  This experiment also did not incorporate movement relative to the ECI, and so would not engender changes with time in an AST framework.  Additionally, the experimental design is incapable of distinguishing SR and AST because, as described below, the shared kinematics of the two theories ensure null results from the ``stationary'' perspective.

From the perspective of the ``stationary'' observer, both the emitter and the detector are moving at the same velocity.  The ``stationary'' observer would observe a blueshifted emission of light from the emitter and a redshift upon detection.  The blueshift and redshift exactly cancel each other, so that no change in frequency with motion would be detected. The redshifted wavelength at the detector $(\lambda'_{\rm det})$ is calculated by the relativistic Doppler shift equation with the input wavelength being the blueshifted wavelength from the emitter $(\lambda'_{\rm bl})$.  Thus, the wavelength detected is the original wavelength:
\noindent
\begin{equation}
\lambda'_{\rm det} = \lambda'_{\rm bl}
\sqrt{\frac{1 +\frac{v}{c}}{1-\frac{v}{c}}}
= \left(\lambda\sqrt{\frac{1-\frac{v}{c}}{1+\frac{v}{c}}}\,\right)
\sqrt{\frac{1+\frac{v}{c}}{1-\frac{v}{c}}}=\lambda\,.\label{7}
\end{equation}
For similar reasons, a test of light speed anisotropy that analyzed frequency changes in a one-way light path,\cite{53} would be unable to distinguish SR and AST.

Additional optical experiments have been recognized to produce equivalent results for SR and ALT: an interferometer experiment with a three-angle closed light path through vacuum and glass;\cite{54,55} a one-way interferometer experiment through air and
water;\cite{56,57} and an experiment on the relative frequency of two ammonia masers arrayed in opposite directions.\cite{58}

\section{Derivation of the Relativistic Time Offset Equation}\label{sec3}
In 1910, Comstock published an influential thought experiment that illustrated differential simultaneity.\cite{59}  In the thought experiment, a platform moves uniformly past a ``stationary'' observer with velocity~$v$.  A light flashes from the center of the ``moving'' platform of length~$l$.  An observer on the platform, who considers the platform ``at rest'', observes the light reach both ends of the platform at the same time, with clocks at each end of the platform recording the same clock readings when the light arrives.  In contrast, the ``stationary'' observer observes that the light strikes the back of the platform first because the back of the platform moves towards the light and the front moves away from the light.  Significantly, the ``stationary'' observer would observe that clocks at each end of the ``moving'' platform record the same time when the light strikes, even though the clocks are struck sequentially.  This is because the ``stationary'' observer observes that the back clock is offset forward in time and the front clock is offset backward in time relative to each other.  We will refer to the difference in time between the two clocks (that is observed by the ``stationary'' observer) as the ``relativistic time offset'' (RTO).  For another description of the RTO see Ref.~\refcite{60}.

The RTO can be viewed on Minkowski spacetime diagrams when using the convention of ``intelligent observers'',\cite{61} who take into account the transit times of light signals when determining event timing (Fig.~\ref{fig2}(a)).

Comstock calculated the non-relativistic times that the light would take to reach the back of the platform, $(l/2)/(c+v)$, and the front of the platform, $(l/2)/(c-v)$.  He stated that the difference in the two times, $(l/c)(v/c)/(1-v^2/c^2)$, is the amount by which the clocks disagree for the ``stationary'' observer,\cite{59} i.e.~the RTO.

Comstock's derivation does not incorporate the fact that the ``stationary'' observer would view a length-contracted platform on which the clocks run slower.  These relativistic effects need to be incorporated, as they affect the clock readings on the ``moving'' platform that a ``stationary'' observer would observe.  Here, we derive the RTO equation taking the relativistic effects into consideration.

As viewed by the ``stationary'' observer, the time for the light to travel from the center to the back of the platform in the ``moving'' reference frame is the time transformation equation (\ref{3}) in which the input for $t$ is the time that the light takes to cross the length-contracted platform, $(l(1-v^2/c^2)^{0.5}/2)/(c+v)$:
\begin{equation}
t'_{\rm back}=\frac{l\sqrt{1-\frac{v^2}{c^2}}}{2(c+v)}
\sqrt{1-\frac{v^2}{c^2}}=\frac{l\!\left(1-\frac{v}{c}\right)}{2c}\,.\label{8}
\end{equation}
The corresponding time for the light to reach the front clock is:
\begin{equation}
t'_{\rm front}=\frac{l\sqrt{1-\frac{v^2}{c^2}}}{2(c-v)}
\sqrt{1-\frac{v^2}{c^2}}=\frac{l\!\left(1+\frac{v}{c}\right)}{2c}\,.\label{9}
\end{equation}
The RTO is obtained by subtracting $t'_{\rm back}$ from $t'_{\rm front}$:
\begin{equation}
{\rm RTO} =\frac{vl}{c^2}\,.\label{10}
\end{equation}
It is not unexpected that the derivation of the RTO gives $vl/c^2$, as this is equivalent to the $vx/c^2$ term in Eq.~(\ref{2}), which alters clock times with distance to allow differential simultaneity and isotropic light speeds.

The RTO equation is equivalent to the difference in timing between the anisotropic light propagation time predicted by ALT and the calculated isotropic light speed in the moving frame.  To derive this, consider light sent in the direction of motion: subtracting the isotropic light propagation time, $l/c$, from the anisotropic light propagation time, $l(1+v/c)/c$, gives $vl/c^2$, which is the RTO equation.  This equivalence is expected, as the RTO is implicit in the isotropic light propagation time but is absent from the ALT anisotropic light propagation time, so the difference between the two times is the RTO value.

The RTO equation is also equivalent to the equation for the ``Sagnac correction'' (after substituting for the corresponding angular values).\cite{62}  This allows AST to be compatible with light speed anisotropies observed in satellite communications, as described below.

\section{AST is Compatible with Light Speed Anisotropies Observed in Satellite Communications}\label{sec4}
In 1913, Sagnac discovered that light traveling in a rotating frame strikes a rotating receiver moving towards the light earlier than a rotating receiver moving away from the light. This was interpreted by Sagnac to mean that the light was traveling in a reference frame independent of the rotating frame, thereby providing evidence for an ``ether".\cite{63} In 1914, Witte argued that the observation of anisotropic light speeds within the rotating reference frame did not necessitate the existence of an ``ether" because the motion was rotational and therefore not an IRF in which the second postulate of SR would be expected to hold.\cite{64}

In 1925, Michelson and Gale demonstrated that light speed is anisotropic on the surface of the rotating Earth.\cite{65,66}  An around-the-world Sagnac experiment by Allan {\it et~al.}, 1985, showed that light propagates isotropically in the ECI but not in the rotating earth-centered, earth-fixed (ECEF) reference frame.\cite{67}  The experiment demonstrated that using the distance measured in the rotating ECEF reference frame and the time of one-way light propagation around the Earth produces a calculation of light speed anisotropy.  If instead, the distance was calculated from the non-rotating ECI perspective then the light speed would be isotropic.  Consistent with the Allan {\it et~al.} result, communications between satellites of the global navigation satellite system (GNSS) and receivers linked to atomic clocks on the Earth's surface routinely demonstrate that light propagates isotropically in the ECI but anisotropically if distances are measured in the rotating ECEF reference frame.\cite{16,68}  The Sagnac correction describes the difference between the light propagation times in the ECI versus in rotating reference frames, such as the ECEF.\cite{69}  The application of the Sagnac correction allows a rotating observer (who is measuring distances in the rotating frame) to convert their calculated ``isotropic'' light propagation times to the actual, observed anisotropic values.

Light speed anisotropy is also observed for satellite-to-satellite communications if distances are calculated relative to the satellites.  This is illustrated by the two GRACE satellites that follow each other in the same orbit at a distance of $\sim$\,220~km.\cite{70}  The GRACE satellites use USO (ultrastable oscillators) to send micro\-wave signals with encoded time information between the
satellites.\cite{71}  The comparison of the time-stamped signal when it is received relative to the USO time on the receiving satellite is used to calculate the time of flight of the microwave signal.  This time difference is used to calculate the distance between the satellites.\cite{71}  The distance is also independently determined by geodetic-quality GPS receivers on the GRACE
satellites.\cite{71,72}  Viewed from the ECI perspective, the leading satellite moves away from the light signal sent by the lagging satellite, while the lagging satellite moves towards the light signal sent by the leading satellite.  Calculating the speed of light based on the instantaneous distance between the satellites and the observed light propagation times produces anisotropic values that are described by the Sagnac correction.\cite{73}

Light speed anisotropy has also been observed for motion that has both rotational and non-rotational components.
Vessot {\it et~al.} calculated the speed of light as isotropic in the ECI reference frame for communications between Earth-based stations and a rocket.\cite{19,20}  In the Vessot {\it et~al.} experiment, the timing of light signals were made from the ``moving'' Earth-based stations and the rocket, using their ``moving'' clocks.  Within the limits of the experiment, the results support isotropic light speeds when calculated in the ECI reference frame, providing differences relative to the isotropic light speed of
$1.9\times 10^{-8}$; $-3.2\times 10^{-9}$ and $-5.6\times 10^{-8}$ for different parts of the trajectory.  The small differences between the observed one-way light speeds and the expected isotropic light speed when calculated in the ECI, implies that if the calculations were made from the rotating ECEF, the light speed would be found to be anisotropic in the ECEF reference frame.

The experiments described above utilized observations made in the ``moving'' reference frame because both the Earth's surface and the satellites are ``moving'' relative to the ECI, which is considered the isotropic ``stationary'' reference frame for the calculations of light speed.   When light signals are sent in IRFs, SR predicts that the RTO would affect clock timing so that the ``moving'' observers would calculate isotropic light speeds of $c$.  The satellite data indicates that there is no appreciable RTO affecting the ``moving'' clocks.  Within experimental limits, the light speed anisotropy calculated by the ``moving'' observers matches that predicted by the Sagnac correction, which does not incorporate a RTO.\cite{67}  Even a significant partial RTO would manifest itself in the GNSS communications that transfer high resolution universal coordinate time (UTC) information.\cite{74}  Therefore, one can infer that clocks do not experience a detectable RTO in rotating reference frames.  This observation is consistent with AST, which predicts the absence of a RTO for both IRFs and non-IRFs.

Both SR and AST are considered to be compatible with light speed anisotropy in rotating reference frames,\cite{75,76,77,78,79} which makes it difficult to use measurements of light speed in rotating frames to distinguish the two theories. In contrast, SR predicts light speed isotropy for measurements made within an IRF, while AST predicts light speed anisotropy.  Critically, this distinction applies only to IRFs that are not considered to be a PRF in the AST framework, such as the ECI or HCI.  Therefore, experiments to distinguish the two transformations must utilize IRFs that are not AST-defined PRFs.  Additionally, measurements of distance and time must be carried out by ``moving'' observers in the IRF.  Below, we describe an experimental strategy that analyzes light speed within a linear, constant velocity IRF to directly measure the RTO, and thereby distinguish the two theories.

\section{A Strategy to Directly Test Differential Simultaneity}\label{sec5}
\begin{figure}[t]
\vspace*{-2pt}
\hfil
\begin{minipage}[b]{5.3cm}
\myfig{width=5.3cm}{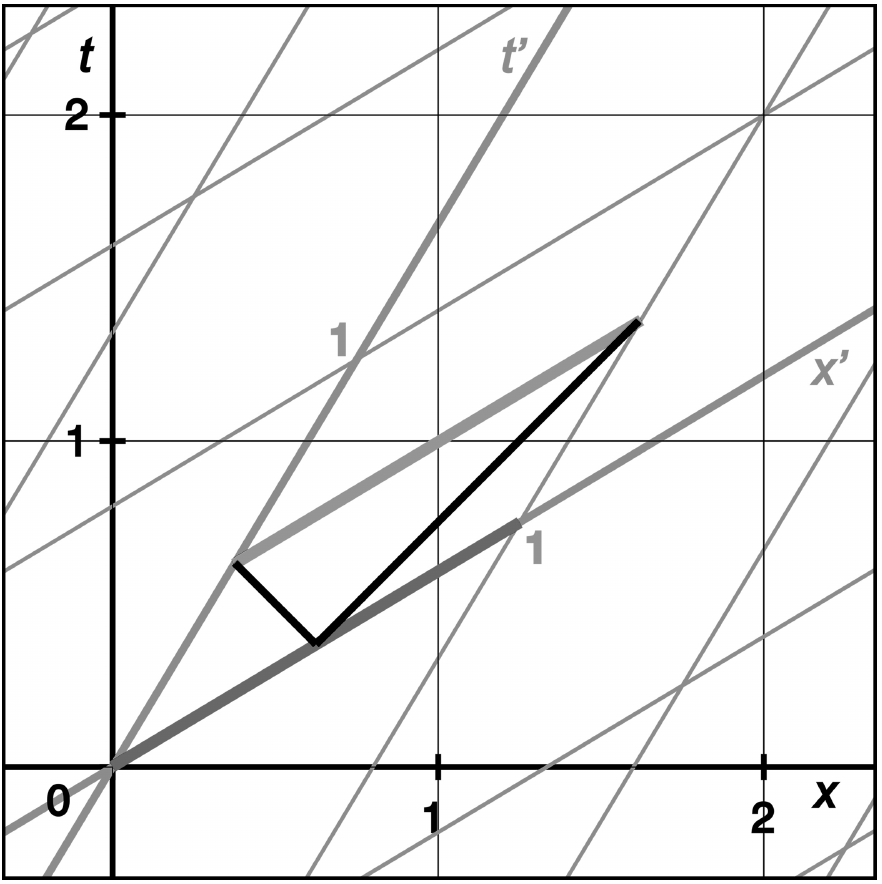}
\vspace*{4pt}
\centerline{\footnotesize (a)}
\end{minipage}
\hfil
\begin{minipage}[b]{5.3cm}
\myfig{width=5.3cm}{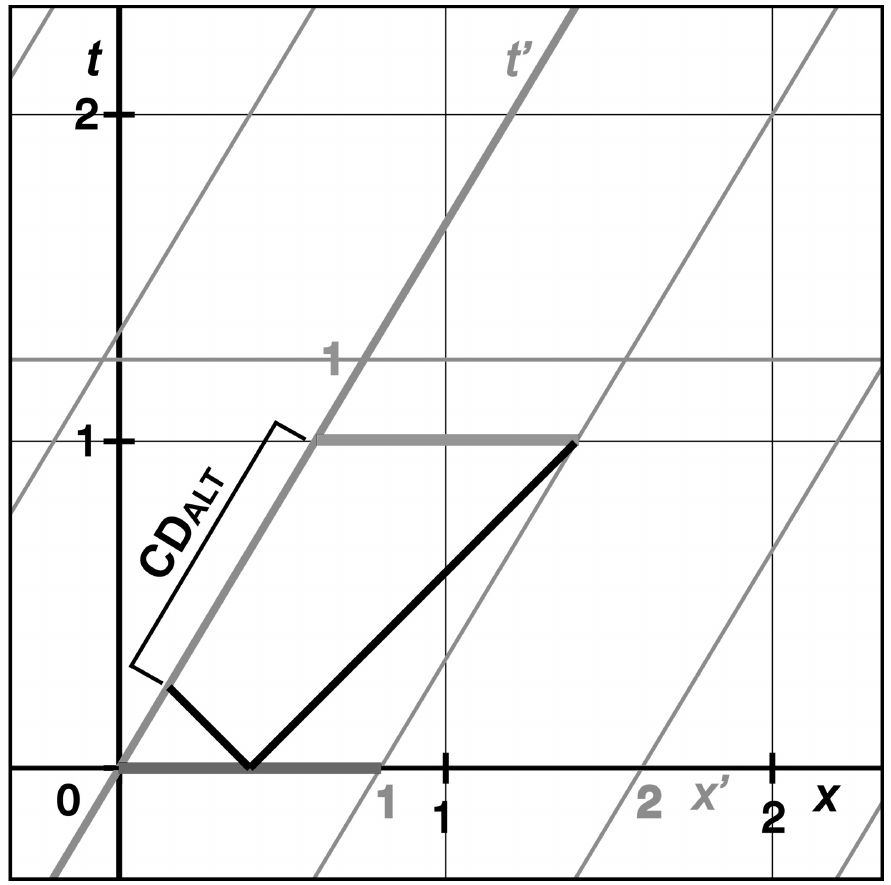}
\vspace*{4pt}
\centerline{\footnotesize (b)}
\end{minipage}\caption{(a)~Minkowski and (b)~ALT spacetime diagrams for $v = 0.6c$ illustrating clock differences.  For both diagrams, the initial position of a vehicle 1 light-fs in length is marked by a thick dark-gray line and the final position of the vehicle is marked by a thick light-gray line.  At $t' = 0$, light beams (black lines) are sent from the middle of the vehicle towards the two ends of the vehicle.  In (a), the ``moving'' observer observes that the two light beams strike the ends of the vehicle simultaneously, producing a null $CD$ value that implies a RTO.  In (b), the moving observer observes a $CD_{\rm ALT}$ of 0.6~fs that implies no RTO.}
\label{fig3}
\vspace*{5pt}
\end{figure}

Comstock discussed his thought experiment from the perspective of the ``stationary'' observer.  However, by utilizing ``moving'' observers to measure the clock readings in the ``moving'' reference frame, a similar experimental strategy can be used
to distinguish SR and AST.  The SR ``moving'' observer will observe the light strike the front and back clocks at identical clock
readings of $(l/2)/c$ (Fig.~\ref{fig3}(a)). The ALT moving observer will observe the light strike the back and front clocks with times described by Eqs.~(\ref{8}) and (\ref{9}), respectively.  The difference in the two times is equivalent to the RTO, $vl/c^2$, and will be referred to as the ``clock difference'' for ALT, $CD_{\rm ALT}$ (Fig.~\ref{fig3}(b)).

\begin{figure}[t]
\myfig{width=9.0cm}{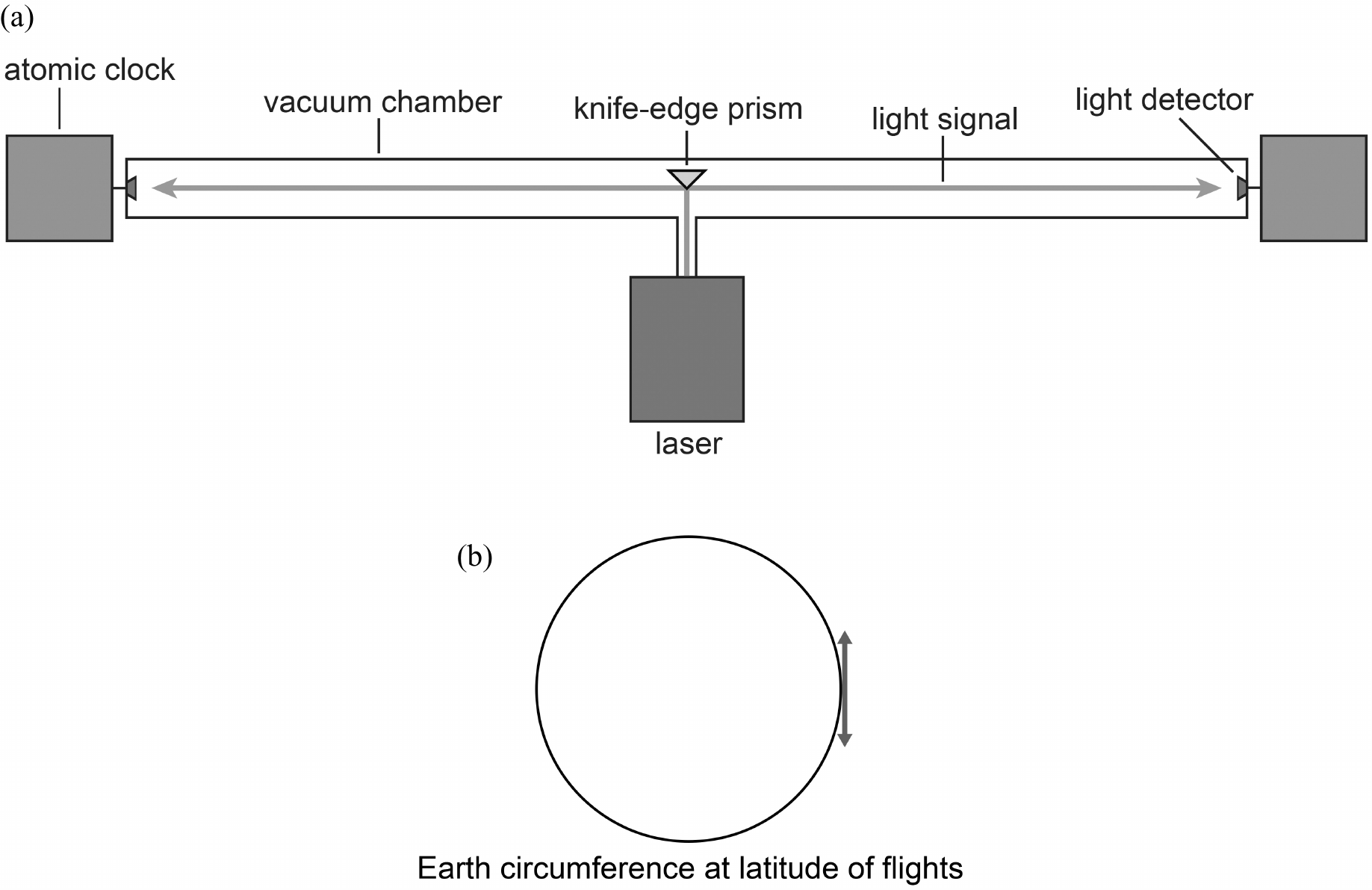}
\caption{(a)~Diagram of the $CD$ measurement apparatus.  (b)~Diagram of the Earth's circumference at the latitude of the flight.  Flight trajectories are shown with arrows.}
\label{fig4}
\end{figure}

An experimental strategy that would be capable of measuring $CD$ values is described here.  The components include an elongated chamber under vacuum with light detectors at each end, and a laser capable of femtosecond pulses aligned at a right angle to the center of the chamber (Fig.~\ref{fig4}(a)).  Across from the laser is a knife-edge prism that is mounted with a piezo-based picomotor that allows fine longitudinal movements.  Light pulses from the laser are split by the knife-edge prism and directed to each end of the vacuum chamber where detectors receive the light signals and atomic optical clocks record the time.  The difference in the time readings is the $CD$.

To synchronize the two clocks, the apparatus is placed in a north--south orientation on the Earth's surface.  In this orientation, both clocks are at the same $x$-axis position relative to the rotational motion of the Earth.  The times of the clocks are synchronized in the north--south orientation so that when the light hits the two detectors, they give identical clock readings.  The apparatus is then shifted to a west--east orientation, which separates the clocks in the direction of motion relative to the ECI.  For SR, this has the effect of shifting the two clocks to different times relative to a ``stationary'' ECI observer, i.e.~introducing the RTO (Fig.~\ref{fig2}(a)).  In the SR framework, the RTO ensures that no $CD$ value is observed in the ``moving'' reference frame, which implies differential simultaneity.  AST does not have a RTO, and so a stationary ECI observer would observe that the two clocks maintain their synchrony after the shift (Fig.~\ref{fig2}(b)).  A $CD$ value equivalent to $CD_{\rm AST}$ (which is the $CD_{\rm ALT}$ value with velocity calculated relative to an AST-defined PRF) signifies no RTO, which implies absolute simultaneity.

The synchronization scheme utilizes synchronization at the same location (relative\- to the direction of motion) followed by equivalent slow clock transport to spatially separate the clocks in the direction of motion.  While it may be argued that this synchronization assumes the direction of motion, this does not affect the ability of the experiment to distinguish between SR and AST.  SR predicts a null $CD$ after synchronization and movement in any direction, while AST predicts an exact $CD_{\rm AST}$ value for this experimental strategy.

The experiment can be undertaken in an IRF with a linear, constant velocity relative to the ECI.  For example, an airplane in eastward and westward trajectories that are linear relative to the ECI, as shown in Fig.~\ref{fig4}(b).  For SR, flights in both directions are predicted to produce null $CD$ values.  For AST, the eastward flight would have a greater velocity relative to the ECI.  Consider an airplane traveling at 250~m/s at the latitude of Boulder, CO, USA (which has an angular velocity of 356~m/s due to the Earth's rotation).  The velocity relative to the ECI would be $\sim$\,606~m/s when flying eastward and $\sim$\,106~m/s when flying westward.  The $CD_{\rm AST}$ for a 10~m-long apparatus on the eastward flight would be 67.4~fs, while the $CD_{\rm AST}$ for the westward flight would be 11.8~fs.

The experimental strategy described above is currently not technically feasible using velocities available with airplanes because it requires distinguishing timing differences in the range of 10--50~fs.  Optical atomic clocks operate at 0.9--2.3~fs per frequency cycle, which would, in principle, be sufficient to detect such timing\break differences.\cite{80,81,82,83}  However, it is currently not possible to count individual frequency cycles with these optical clocks.\cite{84}  Time-to-digital converters can measure 0.1~pico\-second (ps) time intervals,\cite{85} but this does not provide sufficient resolution.

The experimental strategy is technically feasible if carried out in a space-based setting where greater velocities are possible.  A suitable IRF could be a spacecraft traveling linearly near a planet in a direction opposite to the planet's orbital motion.  For example, a spacecraft that was traveling at 12~km/s (in the HCI) in a direction opposite to the Earth's orbital motion of 30~km/s would have a velocity of 42~km/s relative to the ECI.  In this setting, the $CD_{\rm AST}$ for a 10~m-long experimental apparatus is predicted to be 4.67~ps, which is detectable with current time-to-digital converters.  Notably, when the spacecraft is distant from planetary SOIs, the $CD_{\rm AST}$ is predicted to be 1.34 ps based on a velocity of 12~km/s relative to the HCI.  These timing differences can be distinguished using current time-to-digital converters.

\section{Controls and Statistical Tests for the Experimental Strategy}\label{sec6}
Control tests can be used to assess random and systematic errors.  Random and non-directional errors can be identified by determining if the clock readouts accurately reflect the time of light propagation.  This can be checked by moving the knife-edge prism closer to one end of the apparatus so that the difference in distance that the light travels in each arm of the apparatus should produce a defined, positive reading.  For example, moving the knife-edge prism 20~$\mu$m from the center with the piezo-based picomotor would be expected to produce a timing difference of 67~fs in a 10~m-long apparatus when in a north--south orientation.

Tests to determine if there are systemic or directional errors can be accomplished by altering the orientation of the apparatus.  Rotating the apparatus $180^{\circ}$ when it is perpendicular to the direction of motion should maintain clock synchronization if there are no directional errors.  Rotation of the apparatus $180^{\circ}$ in the direction of motion would be expected to give the same $CD$ value but with an opposite sign for AST, or maintain the null $CD$ value for SR.  To ensure that the alignment of the apparatus towards either the front or rear of the vehicle does not affect the clock readings, the eastward or westward trajectory can be reversed, which should alter the $CD$ value based on velocity relative to the ECI for AST, or maintain the null $CD$ value for SR.  These tests can ensure that systematic errors arising from the apparatus are not interpreted as valid results.

The speed of the apparatus is a variable that affects the predicted $CD_{\rm AST}$.  $CD_{\rm AST}$ is predicted to increase by 0.111265 fs for every m/s increase in velocity relative to the ECI for a 10~m-long apparatus.  The speed of the apparatus during the data collection period should therefore be recorded and used to calculate the predicted $CD_{\rm AST}$ values.

Experimental tests of SR often use test theories, such as Mansouri and Sexl (MS), to assess LT parameters.\cite{86}  The MS test theory equations are:
\begin{eqnarray}
t' &=& at + \epsilon x'\,,\label{11}\\[6pt]
x' &=& b(x - vt)\,,\label{12}\\[6pt]
y' &=& dy\,,\label{13}\\[6pt]
z' &=& dz\,.\label{14}
\end{eqnarray}
The parameters that are experimentally analyzed, $a(v)$, $b(v)$ and $d(v)$, are the same for SR and ALT  (when calculated from the ``stationary'' perspective): the time dilation factor $a(v) = (1-v^2/c^2)^{0.5}$; the $x$-axis length contraction factor
$b(v) = 1/(1-v^2/c^2)^{0.5}$; and the $z$-axis and $y$-axis length contraction factor
$d(v) = 1$.\cite{10}
The relevant term in the MS test theory that distinguishes the two transformations is $\epsilon(v)$, which depends on the method of clock synchronization.  $\epsilon(v)$ is $-v/c^2$ for SR and~0 for ALT.\cite{10}  An experimentally-determined $CD$ can be used to determine the corresponding value of $\epsilon(v)$, as the two functions are related as described below.

Using Eqs.~(\ref{11}) and (\ref{12}) with the SR/ALT relativistic values of $a(v)$ and $b(v)$ to produce spacetime diagrams shows that the value of $\epsilon(v)$ determines the slope of the $x'$ axis (Fig.~\ref{fig3}).  In contrast, the slope of the $t'$ axis is not affected by the value of $\epsilon(v)$.  $\epsilon(v)$ can be described in terms of $v$ and the slope of the $x'$ axis, $m$, by solving the relativistic version of Eq.~(\ref{11}) with $t'=0$ (to represent the $x'$ axis), and substituting $x'$ with the relativistic version of Eq.~(\ref{12}).  Because $t'$ is set to~0, $t/x$ is equivalent to $m$.  $\epsilon(v)$ in terms of $m$ and $v$ is:
\noindent
\begin{equation}
\epsilon = \frac{-m\!\left(1 - \frac{v^2}{c^2}\right)}{(1 - mv)}\,.\label{15}
\end{equation}
The $CD$ value also determines the slope of the $x'$ axis on a spacetime diagram.  The expression of $CD$ in terms of the slope of the $x'$ axis can be determined by geometrically solving for $CD$ (Fig.~\ref{fig3}(b)).  $CD$ in terms of $m$, $v$ and $l$ is:
\noindent
\begin{equation}
CD = \frac{l\!\left(\frac{v}{c^2}-m\right)}{(1 - mv)}\,.\label{16}
\end{equation}
Subtracting the description of $CD/l$ in terms of $m$ and $v$, shown in Eq.~(\ref{16}), from the description of $\epsilon(v)$ in terms of $m$ and $v$, shown in Eq.~(\ref{15}), gives $-v/c^2$.  Therefore, $\epsilon(v)$ is related to $CD$ through the following equation:
\noindent
\begin{equation}
\epsilon =  - \frac{v}{c^2}+\frac{CD}{l}\,.\label{17}
\end{equation}
Equation~(\ref{17}) allows the value of $\epsilon(v)$ to be calculated from the experimentally-determined $CD$ value.  $z$-statistics can be used to determine the confidence intervals of the calculated $\epsilon(v)$ values to determine the extent to which the data matches the predictions of SR $(-v/c^2)$ or AST (0).

\section{Conclusions}\label{sec7}
This work highlights that AST is the only viable alternate to SR.  Malykin and Vargas have shown that ALT is the only alternate LT that is compatible with classical tests of SR.\cite{1,2}  We provide proofs that show that AST is compatible with optical resonator experiments and experiments analyzing frequency changes in one-way light pathways.  We also summarize the literature to show that other light speed isotropy/anisotropy experiments are compatible with both AST and SR.  Therefore, current light speed experiments fail to distinguish AST from SR.

It has been proposed that ALT is an alternative form of the LT because clocks can be resynchronized to convert ALT to the LT.\cite{9}  We point out that this resynchronization of clocks would abolish the absolute simultaneity characteristic of ALT.  Further, we note that this resynchronization of clocks cannot be implemented in an absolute simultaneity framework because it is not possible to resynchronize clocks so that a given clock will simultaneously match the different times expected by different IRF observers.   It has also been proposed that resynchronization would allow the PRF in ALT to switch seamlessly between different IRFs, so that previous PRFs would become non-PRFs.\cite{12}  While this scenario would allow ALT to match all experimental data (by choosing the PRF to match the observed data), it lacks a plausible mechanism that would implement the resynchronization of clocks in a real-world setting.  In the absence of resynchronization, ALT is only compatible with experimental data if the PRFs proposed by ALT are locally associated with gravitational centers, similar to the concept of SOIs in astrodynamics.

The RTO is an integral aspect of differential simultaneity, but is absent in absolute simultaneity.  Our derivation of the RTO shows that it is identical to the Sagnac correction, which describes the extent of light speed anisotropy calculated in a rotating reference frame relative to the isotropic propagation of the light in the ECI.  The extensive use of the Sagnac correction in satellite communications highlights that light speed is isotropic in the ECI but not in the rotating reference frame.  This data is compatible with AST because it shows that there is no appreciable RTO affecting the clock times of the ``moving'' observers (satellites and ground-based receivers).

AST predicts light speed isotropy within PRFs that are locally associated with centers of gravitational mass, such as the ECI, other planetary non-rotating IRFs and the HCI.  The proposed experimental strategy assesses light speed isotropy within linear, constant velocity IRFs that are not AST-defined PRFs.  Other potential light speed tests that would be capable of distinguishing SR and AST have been proposed.\cite{87,88,89}  A major advantage of the experimental strategy described here is that it provides a direct readout of the RTO, and therefore a direct measure of differential versus absolute simultaneity.  The proposed experimental strategy is currently technically feasible only in a space-based experiment.  However, we anticipate that continuing technical advances will make this and similar experimental strategies more accessible, so that direct experimental tests of the nature of simultaneity can be accomplished in the near future.

\section*{Acknowledgment}
We thank J. Reeves and the University of Georgia Statistical Consulting Center for advice on statistical analysis;
and J. L. Wilmoth and B. E. Kipreos for helpful discussions.

\end{document}